\documentclass[aps,floats,superscriptaddress]{revtex4}
\usepackage{amsmath}
\usepackage{graphicx}
\allowdisplaybreaks[1] 

\def\GroupeEquations#1{\begin{subequations}  #1  \end{subequations}}
\def\moy#1{\left\langle #1 \right\rangle}
\def\dmoy#1{\left\langle \! \moy{#1} \! \right\rangle}
\def\dkmoy#1{\left\langle \! \moy{#1} \! \right\rangle_{{\cal K}}}
\def\kmoy#1{\moy{#1}_{{\cal K}}}

\def\Im{\hbox{Im}}
\def\Tr{\text{Tr}}

\def\sgn{\text{sgn}\,}

\def\fig#1#2{\includegraphics[height=#1]{#2}}
\def\figx#1#2{\includegraphics[width=#1]{#2}}

\newcommand{\empile}[2]{\genfrac{}{}{0pt}{}{#1}{#2}}

\def\matrice#1{{\begin{pmatrix}#1\end{pmatrix}}}

\def\SigmaSkel{{ \Sigma \kern -5.5pt \raise 1pt \hbox{/}}}
\def\Mdot{M_{\text{dot}}}

\begin{document}

\title{On the perturbative expansion of the magnetization \break in the
out-of-equilibrium Kondo model.}

\author{O. Parcollet}\affiliation{Center for Materials Theory,
Department of Physics and Astronomy, Rutgers University, Piscataway, NJ
08854, USA}  
\affiliation{Service de Physique Th{\'e}orique, CEA Saclay, 91191 Gif-Sur-Yvette, FRANCE}

\author{C. Hooley}
\affiliation{Center for Materials Theory,
Department of Physics and Astronomy, Rutgers University, Piscataway, NJ
08854, USA}  
\affiliation{School of Physics and Astronomy, Birmingham University,
Edgbaston, Birmingham B15 2TT, UK}

\date{May 15th, 2002}

\begin{abstract}
\ \par
This paper is concerned with the out-of-equilibrium two-lead Kondo model, considered
as a model of a quantum dot in the Kondo regime.  We revisit the perturbative expansion
of the dot's magnetization, and conclude that, even at order 0 in the Kondo interactions,
the magnetization is not given by the usual equilibrium result.  We use
the Schwinger-Keldysh method to derive a Dyson equation describing the steady state
induced by the voltage between the two leads, and thus present the correct procedure
for calculating perturbative expansions of steady-state properties of
the system.  
\end{abstract}
\maketitle

\section{Introduction}
In recent years, much experimental and theoretical
work has been devoted to exploring the properties of so-called `quantum
dots' \cite{GlazmanReview}.
These are mesoscopic devices in which a `dot' containing a small number of electrons is isolated   
from two macroscopic leads (denoted `left' and `right') by potential barriers,  
through which tunneling processes take place. Experimentally, these systems are
small devices fabricated using a two-dimensional electron gas 
\cite{twodeg1,twodeg2,twodeg3,twodeg4} or carbon nanotubes \cite{nanotube1}.
The number of electrons on the dot, $N$, is 
controlled by a gate voltage $V_{\rm g}$.  From the experimental point of view, 
one is primarily interested in the current flowing through the 
dot as a function of $V_{\rm g}$ and of the potential difference $V$ between the two leads.
When tunneling is weak enough, the Coulomb blockade phenomenon appears \cite{GlazmanReview}: 
the conductance through the dot is essentially zero except in the vicinity of certain special values of $V_{\rm g}$,
where the energy difference between the ground states of the dot with $N$ and $N+1$ electrons vanishes.
At these points, conductance peaks are observed.

A simple model for such a system is the Anderson model, where the localized level represents the dot and
the hopping term describes its hybridization with the leads.
In this paper, we shall restrict ourselves to the regime where the occupation of the dot is not fluctuating,
and where na{\"\i}ve application of Coulomb blockade ideas would predict a strongly suppressed conductance.
However, when $N$ is odd there is exactly one unpaired spin, which is coupled to the leads
via a Kondo interaction.
In this case, in the linear response regime, spin physics opens up a new transmission channel via the Kondo effect.
Although the occupancy of the dot remains fixed, spin-flip interactions
permit the formation of strong dot-lead hybridization for
temperatures $T<T_K$, where
$T_K$ is the Kondo temperature.  As the temperature approaches zero,
this leads to unitary limit  conductance ($G = G_0 \equiv 2e^2/h$)
via a sharp resonance
at the Fermi surface---the Abrikosov-Suhl resonance.  This effect was
predicted in the context of  quantum dots
fourteen years ago \cite{GlazmanRaikh,NgLee}, and was recently observed in a
series of experiments \cite{twodeg1,twodeg2,twodeg3,twodeg4,nanotube1}.

By contrast, in the large-voltage regime, full non-equilibrium calculations are required, and much
less is known than close to equilibrium.
The out-of-equilibrium Anderson and Kondo models have been studied by various methods in the last decade.
Much of this work \cite{nca1,nca2,AchimPRL}
has concentrated on the non-crossing approximation (NCA) approach, which adopts a slave-boson
description of the problem, and then renders it tractable by neglecting certain vertex corrections.
This yields a picture in which the Kondo resonance in the density of states
is both split and broadened as the voltage $V$
between the leads is increased.
Recent work \cite{SaleurDotExact} has explored a new approach, where one attempts to use the
Bethe Ansatz results for the Anderson model to construct a Landauer-type picture of transport
through the dot.  This approach also involves approximation, when one comes to construct the
`in' and `out' scattering states from the dressed excitations of the model.
Another thread \cite{pertU1,pertU2} has involved studying the Anderson model via perturbation
theory in the on-site Coulomb repulsion $U$.  While these works provide approximate information
on the behavior of the current-voltage characteristic, they shed little light on
the nature of the many-body state of the system when $V>T_K$.

In particular, a basic question recently debated \cite{WRONGPRL,AchimPRL} is 
whether the Kondo problem has a strong coupling regime at low temperature
and high voltage.
In discussing this point, a previous paper \cite{WRONGPRL} used 
a second order perturbative expression for the
magnetic susceptibility in the out-of-equilibrium steady state induced by $V$.
This putative result was however incorrect, even at order 0 in $J$, the strength of the
Kondo couplings; in this note, we correct this result at order 0, discuss the related physics and 
present a systematic method for calculating higher-order corrections in $J$. 
Our main conclusion is that, even at order 0 in $J$, the Keldysh function of the spin deviates
from its equilibrium value and therefore the steady state magnetization of the dot is not given by 
$M_{\rm eq} \equiv \frac{1}{2} \tanh (B/2T)$. Rather, it must be computed by solving a
transport equation.
We emphasize that this issue is not directly related to the so called ``decoherence time'' \cite{AchimPRL}
but is a basic point about perturbation theory to be addressed before discussing out-of-equilibrium RG equations
and the existence of a strong coupling regime at large voltage. In particular, it has ramifications for other
physical 
quantities, e.g.\ the current and the current-current correlation
function in a magnetic field (see section \ref{sub.derivationEq4}).

The paper is organized as follows. In Section \ref{s:model}, we present the model, our main result 
(the magnetization at order 0 given by Eq.~\ref{MAINRESULT}), 
 and the associated physical discussion.
In Section \ref{PerturbativeExpansion}, we present a detailed pedagogical derivation of 
(\ref{MAINRESULT}) using the Keldysh method.
Finally, in Appendix \ref{App.pointsplitting}, we present more details about the
computation presented in ref.~\cite{WRONGPRL}, and explain why it was incorrect.

\section{Results and discussion}\label{s:model}

Our starting point is the 
Hamiltonian of the two-lead Kondo model. For a discussion of the modeling of the quantum dot, and
for a derivation of this model from the Anderson model via an out-of-equilibrium
Schrieffer-Wolff transformation, we refer the reader to the literature, in
particular \cite{GlazmanKaminskyRG} and references therein. The Hamiltonian is:
\begin{align}\label{ham}
H  = \, & \sum_{\alpha {\bf k}\sigma} 
\bigl (\varepsilon_{{\bf k}} - \mu_{\alpha} \bigr) c^\dag_{\alpha
{\bf k} \sigma} c_{\alpha {\bf k} \sigma} + H_{\rm refl} + H_{\rm
trans} - B_{s} S_{z} - B_{c} 
\sum_{\alpha{\bf k}\sigma}\sigma  c^\dag_{\alpha{\bf k}\sigma} c_{\alpha{\bf k}\sigma},
 \\
H_{\rm refl}  = & \,J_{R}
\sum_{{\bf k},{\bf k}',\sigma,\sigma'}
\left(
c^\dag_{R{\bf k}\sigma} {\vec \sigma}_{\sigma \sigma'}
 c_{R {\bf k}' \sigma'} \right)
\cdot {\vec S} \,\, + \,\, (R \leftrightarrow  L), \nonumber \\
H_{\rm trans}  = & \,
J_{RL}
\sum_{{\bf k},{\bf k}' , \sigma,\sigma'}
 \left(
c^\dag_{R{\bf k}\sigma} {\vec \sigma}_{\sigma \sigma'} c_{L {\bf k}' \sigma'} \right)
\cdot {\vec S} \,\, + \,\, (R \leftrightarrow L), \nonumber
\end{align}
where $c^\dag_{\alpha {\bf k} \sigma}$ creates an electron in lead
$\alpha \in \left\{L,R\right\}$ with momentum ${\bf k}$ and spin
$\sigma$, and $J_L$, $J_R$ and $J_{LR}=(J_{RL})^{*}$
are Kondo coupling constants between the electrons and the spin of the dot $\vec{S}$. 
The first term in $H$ describes the electrons in the leads,
with $\varepsilon_{{\bf k}}$ being the bare energy of an electron of momentum
${\bf k}$ at zero voltage (the same for each lead) 
and $\mu_{\alpha }$ is the  potential in the lead $\alpha$.
Each lead consists of a free electron gas with a density of states
$\rho (\epsilon)$ of bandwidth $D$: ultimately, we will be interested in the result in the large bandwidth
($D\rightarrow \infty$) limit, but the computations are first performed
for finite $D$. We will make the physical
assumption that the leads are in thermal equilibrium at a temperature $T$. 
The voltage is applied by taking the
chemical potentials of the two leads to be different, $\mu_{L}-\mu_{R}=V$.
$H_{\rm refl}$ describes regular Kondo processes, where an electron from a given
lead is spin-flip scattered back into the {\em same\/} lead; $H_{\rm trans}$
describes `spin-flip cotunneling', where an electron from one lead is
spin-flip scattered into the {\em other\/} lead.
If the model (\ref{ham}) is derived from the Anderson model,
one finds that $J_{R}J_{L}= |J_{RL}|^{2}$. In this paper, however, we  relax this relation between the coupling
constants and treat them as independent parameters.
Finally, the last terms represent the coupling to the magnetic field. We
allow two different magnetic fields, $B_{s}$ for the spin and $B_{c}$
for the lead electrons; this permits the calculation of the total and
local spin susceptibilities within the same computation.

In this paper, we shall be interested in the values taken by
the following quantities in the non-equilibrium steady state induced by the voltage $V$:
the dot magnetization $M_{\rm dot}$, the
magnetization of the leads $M_{\rm leads}$, the total magnetization $M_{\rm tot}$, and the
total and local susceptibilities. These are given by:
\GroupeEquations{\label{DefM}
\begin{align}
M_{\rm dot} (B_{s},B_{c}) &= \moy{S_{z}}, \label{DefMdot}
\\
 M_{\rm leads} (B_{s},B_{c}) &=  \moy{\sum_{\alpha{\bf
k}\sigma}\sigma  c^\dag_{\alpha{\bf k}\sigma} c_{\alpha{\bf k}\sigma}}, \label{DefMleads}
\\
M_{\rm tot} (B)&=M_{\rm dot} (B,B) + M_{\rm leads} (B,B)- M_{\rm Pauli},
\\
\chi_{\rm tot}&=\left. \frac{\partial M_{\rm tot} (B)}{\partial B} \right\vert_{B=0},
\\
\chi_{\rm loc}& =\left. \frac{\partial M_{\rm dot} (B_{s},B_{c}=0)}{\partial B_{s}} \right\vert_{B_{s}=0},
\end{align}
}
where angle-brackets $\langle \ldots \rangle$ denote an expectation value taken
in the steady (i.e.\ long-time) state of the system.  $M_{\rm Pauli}$ is simply
the Pauli paramagnetic contribution from the lead electrons which would be present even in
the absence of the impurity, and which we therefore exclude from $M_{\rm tot}$. 
We consider the perturbative expansions of these steady state quantities; more 
precisely, we define: 
\begin{equation}\label{DefThetas}
J_{R} \equiv  \theta_{R} J \qquad 
J_{L} \equiv  \theta_{L} J \qquad 
J_{RL} \equiv  \theta_{RL} J
\end{equation}
and we let $J$ go to zero while keeping the coefficients $\theta_R$,
$\theta_L$ and $\theta_{RL}$ fixed. In the following, the expression ``order $n$'' refers to the
order $n$ of this expansion in $J$.

Our main result is the  order $0$ term of the perturbative expansion of the magnetization:
\begin{equation}\label{MAINRESULT}
M_{\rm tot} (B) = M_{\rm dot} (B,B_{c}) + O(J) = \frac{1}{2} \tanh \left(\frac{B}{2T} \right) 
\dfrac{\varphi \left(\frac{B}{T} \right)
\left(
1 + \frac{\theta_{R}^{2} + \theta_{L}^{2}}{2\theta_{RL}^{2}}
\right) }
{ \frac{1}{2} 
\Bigl [ \varphi \left(\frac{B+V}{T} \right) +  \varphi \left(\frac{B-V}{T} \right)
\Bigr ] 
 + 
\frac{\theta_{R}^{2} + \theta_{L}^{2}}{2\theta_{RL}^{2}}  \varphi \left(\frac{B}{T} \right)
} + O (J),
\end{equation}
where $\varphi$ is defined by:
\begin{equation}\label{DefinitionPhi}
\varphi (x) \equiv \frac{x}{\tanh \left(\frac{x}{2} \right)}.
\end{equation}
(At this order, $M_{\rm dot} (B,B_{c})$ does not depend on $B_{c}$.)
As a result, the magnetic susceptibility at order 0 is:  
\begin{equation}\label{MAINRESULTSusceptibility}
\chi (T,V) =  \frac{1}{4T}
\dfrac{ 
1 + \frac{\theta_{R}^{2} + \theta_{L}^{2}}{2\theta_{RL}^{2}}
}
{ \frac{1}{2}\varphi \left(\frac{V}{T} \right)
 + 
\frac{\theta_{R}^{2} + \theta_{L}^{2}}{2\theta_{RL}^{2}}  
} + O (J),
\end{equation}
and in particular for $\vert V \vert \rightarrow \infty$
\begin{equation}
\chi (T,V) \sim \frac{1 +  \frac{\theta_{R}^{2} + \theta_{L}^{2}}{2\theta_{RL}^{2}}}{2 \vert V \vert}.
\label{highVsuscept}
\end{equation}

A striking feature of (\ref{MAINRESULT}) is that the magnetization at order 0 in $J$ is  not given by 
the  equilibrium expression $M_{\rm eq}= \frac{1}{2} \tanh (B/2T)$.
This may seem surprising: since the couplings to the leads are relaxed to zero, 
why should we not find $M_{\rm eq}$, the magnetization of a free spin?
Physically, the answer is that $M_{\rm eq}$ is not the
``magnetization of a free spin'', but rather the magnetization of a
spin weakly coupled to an {\em equilibrium\/} bath.
On the other hand, Eq.~(\ref{MAINRESULT}) gives the magnetization of a
spin weakly coupled, via the Hamiltonian (\ref{ham}), to two
leads with different chemical potentials. It is simply an out-of-equilibrium extension of 
the Curie law. In particular, we emphasize that the finite susceptibility at $T=0$ 
cannot be interpreted as a renormalization (or a screening) of the spin due to the voltage.

At long times, the state of the spin
is completely determined by the
properties of these leads, and hence so is its
distribution function, which describes the population of its two
states as a function of temperature, voltage, and magnetic 
field.  
Since the whole system is not in equilibrium, this steady state is not described by the Gibbs
distribution; in particular, the fluctuation-dissipation theorem (FDT) need not hold.
Hence the magnetization need not be (and is not) $M_{\rm eq}$ at order 0 in $J$. 
Rather it should be computed by solving a transport equation in the steady state regime, i.e.\ a quantum
Boltzmann equation.  (At dominant order this procedure is equivalent to using a semi-classical master
equation; see Appendix \ref{RelBoltzmann}.)

The crucial point is that $J_{R},J_{L},J_{RL}$ are relaxed to zero, {\em assuming that they are
still bigger than the coupling of the spin to any other thermal bath\/}.
If we were to take into account such a coupling (denoted by $J_{0}$) then the
result would cross over to $M_{\rm eq}$ when $J_{R},J_{L},J_{RL} \ll J_{0}$
(all the couplings going to zero while maintaining fixed ratios).
In fact, (\ref{MAINRESULT}) implies that the equilibrium value is only an upper bound: 
$0\leq M_{\rm tot} (J=0) /M_{\rm eq} \leq 1$, which follows from the convexity of $\varphi$. This bound is 
saturated only in equilibrium, i.e.\ for $\theta_{RL}\rightarrow 0$ or $V\rightarrow 0$. 

Moreover, as expected on physical grounds, the non-equilibrium
result is much less universal than the equilibrium one.
In equilibrium, the magnetization of a spin weakly coupled to a bath
depends neither on the properties of the bath except the temperature $T$
nor on the form of the couplings between the spin and the bath. Neither of these
statements holds true for the non-equilibrium Kondo model:
$\Mdot$ (at order 0) depends not only on $T$ but also on the voltage $V$ and on
a ratio involving the parameters $\theta_R$, $\theta_L$, and $\theta_{RL}$.
Thus $\Mdot$ is perturbative in $J$, but not in the three variables $J_{R},J_{L},J_{RL}$,
and in particular it is not analytic in $J_{R},J_{L},J_{RL}$ around
$(0,0,0)$.
One can find such a dependence on the ratio of couplings even in a
simple free model (a single level coupled to two leads), as
illustrated in Appendix \ref{App.Freemodels}.
Furthermore, it should be remembered that we computed $\Mdot$
with free leads: introducing interactions in the
leads would change the function $\varphi$ (the important quantity being the
electron-electron bubble in the leads). For example, we expect a
different result to hold for a spin coupled to Luttinger liquids, even at order 0.

It should be noted that the result (\ref{MAINRESULT}) gives a non-trivial expression
for the magnetization even at zero temperature.
Since $\phi(x) \to \vert x \vert$ as $x \to \infty$, we find that
\begin{equation}
M_{\rm tot} (B,V) \big\vert_{T=0}
= \frac{B}{2} \left( \frac{ 1+ \frac{\theta_{R}^{2} + \theta_{L}^{2}}{2\theta_{RL}^{2}}}{\frac{1}{2}
\left( \vert B+V \vert + \vert B-V \vert \right)
+ \frac{\theta_{R}^{2} + \theta_{L}^{2}}{2\theta_{RL}^{2}} \vert B \vert} \right)
+ O(J).
\end{equation}
We then have two cases (we can take $B,V>0$): for  $B>V>0$, 
we obtain the equilibrium result $M_{\rm tot} (B,V) = \sgn (B)/2 + O (J)$, but for 
$0<B<V$, the magnetization  is a still a function of $B/V$: 
\begin{equation}
M_{\rm tot} \big\vert_{T=0} \approx \frac{B}{2} \left(
\frac{1+\frac{\theta_{R}^{2} + \theta_{L}^{2}}{2\theta_{RL}^{2}}}{\vert V \vert
+\frac{\theta_{R}^{2} + \theta_{L}^{2}}{2\theta_{RL}^{2}} \vert B \vert
} \right),
\end{equation}
in agreement with (\ref{highVsuscept}) in the limit   $\vert V \vert \gg \vert B \vert$. 
Physically, there are two sources of energy available to flip the spin: the thermal fluctuations 
of both baths (represented by $T$) and the fact that an electron can go from L to R 
and give to the spin an energy of order $V$.
If we decrease $B$ from high values ($B \gg T,V$), the spin is locked until $B$ reaches 
the largest of these energy scales. Thus the magnetization at zero temperature is expected to
saturate only for $B>V$. Similarly, the susceptibility
is in general expected to behave as $\chi \sim 1/E$, where $E$ is the
largest energy available to flip the spin.
The fact that our result is still non-trivial at zero temperature implies that 
it could be seen in numerical computations, such as 
the DMRG approach of Cazalilla and Marston \cite{MarsCaz}.

\section{Derivation of Eq. (\ref{MAINRESULT})}
\label{PerturbativeExpansion}

The purpose of this section is to present the 
derivation of Eq.~(\ref{MAINRESULT}) and more generally  the  procedure 
for obtaining the perturbative expansion of physical quantities in the steady state.
It is organized as follows: in paragraph \ref{PrincipleComputation},
we first give an overview of the derivation; full
details are given in the following sections (\ref{KeldyshMethodGal} and
\ref{sub.Perturbationsteadystate}) and in the appendices, including a
presentation of the Keldysh method.

\subsection{Overview}\label{PrincipleComputation}

When doing perturbation theory in the steady state, there are two important small couplings:
$J$, the strength of the
Kondo couplings, and $s$, the small regulator that appears in Green's functions such
as the ``spin'' retarded function (defined in section
\ref{sub.spinpropagators}):
\begin{equation}
R^0_B(\omega) = \frac{1}{\omega-B+is}.
\end{equation}
The scale $s$ should be thought of as being due to the coupling to an
auxiliary thermal bath, which in the physical system would be the substrate.  If the
impurity is coupled to such a bath whose density of states is $\rho_2$ by a coupling
$g$, one finds $s \sim g^2 \rho_2$ (in the large bandwidth limit).  In the physical quantum dot system,
$J \gg s$, meaning that the correct order of limits to take is $s \to 0$ followed by
$J \to 0$, as pointed out in section \ref{s:model}.

However, `straightforward' perturbation theory in $J$ takes the limit
in the opposite order: one first expands in $J $ while keeping $s$ finite,
and only then takes $s \to 0$ term by term in the perturbation series.
In equilibrium, these two limits commute, but out of
equilibrium  they do not. This is explicitly shown in section
\ref{sub.NonCommute}; the out-of-equilibrium Keldysh Green's function is not analytic
around $(s,J)\,{=}\,(0,0)$.  This non-analytic behaviour
is quite generic in situations where an impurity is coupled
to several leads, and may be seen even in a simple free model (see Appendix \ref{App.Freemodels}).
The signature of that non-commutativity is that `straightforward' perturbation theory fails: its terms (starting at
$O(J^2)$) exhibit divergences of a $1/s$ form.  These are similar to the infrared
divergences in equilibrium perturbation theory that signal an incorrect choice of reference
state.  As shown below, $1/s$ divergences at order $J^2$ signal an incorrect choice for the
Keldysh Green's function at order $0$, or equivalently of the
distribution function which describes the non-thermal population of the
two levels of the spin.
Consequently, contrary to the claims of \cite{WRONGPRL}, there is no possibility of
regulating these $1/s$ divergences order by order in $J$.  We discuss the putative regulation
procedure of \cite{WRONGPRL} in Appendix \ref{App.pointsplitting}, and explain why it is
incorrect.

The solution is to begin with the Dyson equation in the steady state, considered as a
functional equation for the full Green's function $G$,
using the skeleton self-energy diagrams (see Appendix \ref{App.DerivationDyson} for further details).
After taking the $s \to 0$ limit
in this equation, a perturbative expansion for $G$ may be inserted into it,
and a solution obtained order by order in $J$.
Solving the Keldysh component of this Dyson equation is equivalent
to solving the quantum Boltzmann equation; as shown
below, the solution is non-thermal even at order $0$ in $J$.
Another procedure for solving the problem of $1/s$ divergences
was proposed in \cite{WingreenSivan} in the context of the $U=\infty$
out-of-equilibrium Anderson model:\ its method was to {\it choose\/} the zeroth-order spin
Keldysh function (see section \ref{sub.spinpropagators}) in order precisely to cancel the
divergences at order $J^2$.  This is completely equivalent to our
approach, as explained in section \ref{sub.derivationEq4}.

Finally, we note that one may give a simple
semiclassical derivation of the out-of-equilibrium result (\ref{MAINRESULT}) based on
a master equation. This derivation is due
to L. Glazman and A. Kaminski \cite{GlazmanKaminskyPrivate}, and is
presented in Appendix \ref{RelBoltzmann}.
The success of such a semiclassical approach (at this
lowest order) is related to the fact that one can compute
the equilibrium magnetization of a free quantum spin using a classical Ising model.
In fact, at this order, the semi-classical master equation is strictly equivalent to 
the Keldysh component of the steady-state Dyson equation, so the
two apparently disparate derivations yield the same result.

\subsection{Technical preliminaries: the Keldysh method}\label{KeldyshMethodGal}

Let us now turn to the technical details. In the following, for simplicity, we will write some equations for a
generic fermionic field $\psi$, which will be specialized afterwards 
to the fields representing the electrons and the spin. 

\subsubsection{Generalities}\label{sub.KeldyshMethod}

The  basic idea of the non-equilibrium Keldysh method
\cite{SchwingerCPT,KeldyshOriginal,RammerSmith} consists in taking the
system at an initial time $t=0$ in an initial state described by a density matrix $\rho_{0}$
and letting the system relax, using the Hamiltonian evolution given by
$H$, to a long-time regime. 
In order to ensure that the system relaxes,
it may be necessary to add some additional coupling terms to the Hamiltonian,
in particular to break conservation laws:\ see the discussion in Appendix
\ref{App.DerivationDyson}.
Depending on the system, the long-time regime
can be an equilibrium state, a
non-equilibrium steady state, a non-time-translation-invariant
steady state, or even an aging regime (in glassy systems). 
In the quantum dot problem, we assume that the system reaches at
finite voltage $V$ a non-equilibrium steady state, in which we want to
compute physical quantities.

A ``Keldysh'' average of any quantity $A$ is defined by 
\begin{equation}\label{DefAvKeldysh}
\kmoy{A (t)} \equiv \moy{e^{i H t} A e^{-iH t}}_{0}\equiv 
\Tr \left( \rho_{0} e^{i H t} A e^{-iH t}\right),
\end{equation} 
where $\moy{\dots}_{0}$ is the average taken using the {\em initial}
density matrix of the system.  The steady-state average is given by:
\[
\moy{A}= \lim_{t\rightarrow \infty }{\kmoy{A (t)}}.
\]
Using the usual representation of the
evolution operator $e^{-i H t}$ as a $T$-ordered exponential in the
interaction picture (and the anti-$T$-ordered one for $e^{i H t}$),
one can obtain an expansion in the coupling constant $J$.
Following the usual conventions, it is convenient to keep track of the two exponentials using
a closed time contour, running from 0 to ${+}\infty$ and back to 0 \cite{KeldyshOriginal}:\
we denote by $+$ the upper contour (from 0 to $\infty$), which arises from
expanding $e^{-i H t}$, and by $-$ the lower contour.
In accordance with this notation, we define the four Green's
functions:
\begin{align}\label{DefGreenFunctions}
G^{++}_{\psi }(t,t') & \equiv  -i \kmoy{ T \psi(t) \psi^\dag(t')},
& G^{-+}_{\psi }(t,t') & \equiv  -i \kmoy{ \psi(t) \psi^\dag(t') }, \nonumber \\
G^{--}_{\psi }(t,t') & \equiv  -i \kmoy{ {\tilde T} \psi(t) \psi^\dag(t') },
&
G^{+-}_{\psi }(t,t') & \equiv  \hphantom{-} i  \kmoy{ \psi^\dag(t') \psi(t) }.
\end{align}
Here $T$ is the time-ordering operator, ${\tilde T}$ the anti-time-ordering
operator, and $\psi$ is any {\em fermionic\/} field.
The two indices of the matrix ${G}$ will be called ``indices in
Keldysh space''. In the following, the equations
(\ref{DefGreenFunctions}) will be summarized with the notation:  
\begin{equation}\label{CompactnotationGreenfunctions}
G_{\psi} (t,t') = -i  \dkmoy{\psi (t) \psi^{\dagger } (t')}
\end{equation}
and $G$ will always denote a $2\times 2$ Keldysh matrix.

The Green's functions (\ref{DefGreenFunctions})
are not independent but can be expressed as
functions of the retarded, advanced and Keldysh Green's functions defined
respectively by \footnote{Throughout this paper, $A$ is the advanced
function and should not be confused with a spectral function.}: 
\begin{align}\label{DefRAK}
R_{\psi }(t,t') & \equiv  -i \, \theta(t-t') 
\kmoy{\left\lbrace \psi(t),\psi^\dag(t') \right\rbrace}, \nonumber \\
A_{\psi }(t,t') & \equiv  i \, \theta(t'-t) 
\kmoy{\left\lbrace \psi(t),\psi^\dag(t') \right\rbrace}, \nonumber \\
K_{\psi }(t,t') & \equiv  -i 
\kmoy{\left[ \psi(t),\psi^\dag(t') \right]},
\end{align}
as is shown from the transformation \cite{RammerSmith}:
\begin{equation}
{ G_{\psi }} \equiv \matrice{G^{++}_{\psi } & G^{+-}_{\psi } \\
G^{-+}_{\psi } & G^{--}_{\psi } } 
\qquad 
{ \widetilde{G}_{\psi }} \equiv  \matrice {R_{\psi } & K_{\psi } \\  0 & A_{\psi } }
\qquad 
{ \widetilde{G}_{\psi }}= \frac{1}{2} \matrice{1 & 1 \\ 1 & -1 }
 { G_{\psi }} \matrice{1 & 1 \\ -1 & 1 }.
\end{equation}
We will denote the first set of Green's functions
(\ref{DefGreenFunctions}) ``the $\pm$ basis'', and the second
(\ref{DefRAK}) the ``Larkin-Ovchinnikov (LO) basis''.

There are further relations between these Green's functions. In
general, $R_{\psi }(t,t') = \left(A_{\psi }(t',t)
\right)^{*}$. Moreover, in equilibrium FDT reads: 
\begin{equation}\label{FDT}
K_{\psi } (\omega ) = h_{\rm eq }  (\omega )\Bigl ( A_{\psi } (\omega ) -
R_{\psi } (\omega ) \Bigr) \qquad 
 h_{\rm eq} (\omega ) \equiv - \tanh \left(\frac{\omega }{2T} \right)
\end{equation}
thus the retarded Green's function is the only remaining
independent Green's function, and so contains all the information about the state of the system.
Out of equilibrium, however, $K_{\psi} (\omega )$ and $R_{\psi} (\omega )$ should be independently determined.

In order to write down the diagrammatic expansion, 
we also define the ``bare'' counterparts of the Green's functions in 
(\ref{DefGreenFunctions}, \ref{DefRAK}) by:
\[
G_{\psi 0}(t,t') \equiv  -i \dmoy{ \psi_{\rm int}(t) \psi_{\rm int}^\dag(t')}_{0},
\] 
where the field $\psi_{\rm int}$ is in the interaction picture.
From the expansions of the evolution operators, we obtain the
diagrammatic expansion, provided that $\moy{\dots}_{0}$
satisfies Wick's theorem. For the problem at hand, we take as the initial condition
the density matrix of the model with $J_{R}=J_{L}=J_{RL}=0$ and for
the spin a finite
$s$ (arising from the coupling of the spin to a thermal bath; see section
\ref{PrincipleComputation} above).
Going to the long time limit, every function becomes a function of the
difference of the times, and transforming to Fourier space we obtain
standard Feynman rules. 

\subsubsection{Lead electrons}\label{sub.leadselectrons}

Let us begin with the lead electrons. We care here only about the
local Green's functions, so we will drop the spatial indices.
We denote by $({\bf G_{c}})_{\alpha\sigma,\beta\sigma'}(t,t')$ the Green's function
describing the creation in lead $\beta$ of an electron with spin
$\sigma'$ at time $t'$, and a corresponding annihilation in lead $\alpha$
of an electron with spin $\sigma$ at time $t$:
\begin{equation}
({\bf G_{c}})_{\alpha\sigma,\beta\sigma'}(t,t') = -i \dkmoy{c_{\alpha\sigma}(t)
c^\dagger_{\beta\sigma'}(t')}.
\end{equation}
(Here, the indices $\alpha,\beta \in \left\{L,R\right\}$, while
$\sigma,\sigma' = +,-$,
and the local
electron operators are defined by $c_{\alpha\sigma} \propto \sum_{\bf k}
c_{\alpha{\bf k}\sigma}$.)
The Green's function ${\bf G_{c}}$ is thus a priori a 
$4\times 4$ matrix (in lead-spin space), whose entries are themselves
$2 \times 2$ Keldysh matrices.  In the following, a bold notation with subscript
$c$ always 
designates such a $4\times 4$ matrix. The bare Green's function ${\bf G_{c0}}$ is
however diagonal. 
The bare (diagonal) density of states is :
\[
(\boldsymbol{\rho_{c0}})_{\alpha \sigma,\beta  \sigma' }= 
\rho (\omega + \sigma B_{c} 
- \mu_{\alpha}) \delta_{\alpha \beta }\delta_{\sigma \sigma'}
\]
(the energy levels are the same is both leads, but shifted by the Zeeman energy and the voltage).
We assume that the baths, being much bigger than the impurity, are
permanently in thermal equilibrium so that the bare electrons'
functions read:
\GroupeEquations{\label{GelectronOrder0}
\begin{align}
{\bf R_{c0}}(\omega ) &=  \int d\epsilon \frac{\boldsymbol{\rho_{c0}} (\epsilon
)}{\omega -\epsilon + i0^{+}},\\
{\bf  K_{c0}} (\omega )&= 2i\pi {\bf h_{c0}} (\omega )
 \boldsymbol{\rho_{c0}} (\omega ),
\\
( {\bf h_{c0}})_{\alpha \sigma,\beta \sigma' } &\equiv 
 h_{\rm eq} \left(\omega - \mu_{\alpha} \right ) \delta_{\alpha \beta }\delta_{\sigma \sigma'},
\end{align}
}
where the first two equations are matricial,
and $\mu_{\alpha }$ is the potential of the lead $\alpha =R,L$. The voltage difference between the leads is
given by $V=\mu_{L}-\mu_{R}$.

\subsubsection{Spin}\label{sub.spinpropagators}

Since the spin operator is not appropriate for diagrammatic
computations (it does not satisfy Wick's theorem), we represent the
spin 1/2 by  three Majorana fermions $\eta^a, a \in \{x,y,z \}$
which satisfy Wick's theorem and the relations
\GroupeEquations{\label{DefMajoranaRepresentation}
\begin{align}
S^a &= - \frac{i}{2} \epsilon^{abc} \eta^b \eta^c, \\
\left( \eta^a \right)^\dagger &= \eta^a,\\
\left\lbrace \eta^a,\eta^b \right\rbrace &= \delta^{ab}.
\end{align}
}
Using (\ref{DefMajoranaRepresentation}),
one can easily show that $\vec{S}$ satisfies the correct commutation
relations and that $\vec{ S}^2 = 3/4$.  Note that this last constraint
is {\it automatically\/} satisfied, unlike in the case of a Dirac fermion
representation, where a Lagrange multiplier would have been required to
fix the magnitude of the spin:\ the Majorana representation
therefore makes the computation simpler.
In this paper, we consider only spin 1/2, but our  computations could
be extended to higher spin, provided that one used another representation
for $\vec{S}$.

Let us now discuss the  propagators ${G}_{ab}$ ($a,b\in
\{x,y,z \}$) of the $\eta$, in the presence 
of a magnetic field $B_{s}$ along the $z$ direction.
The general form is: 
\begin{equation}\label{PropagatorEta1}
{\bf G} = \matrice{G_{xx} & G_{xy} &0 \\
                  G_{yx} & G_{yy} &0 \\
                   0     &  0     &G_{zz} },
\qquad G_{ab} (t,t')\equiv -i \dkmoy{\eta^{a} (t)\eta^{b} (t')},
\end{equation}
where the elements are $2\times 2$ Keldysh matrices.
Indeed $G_{xz}=G_{yz}=0$ by symmetry.  To prove this, note that
the Hamiltonian is invariant under a
$\pi$ rotation around the $z$ axis, which is implemented by
$(\eta^{x},\eta^{y},\eta^{z})\rightarrow(-\eta^{x},-\eta^{y},\eta^{z})$
(and the corresponding rotation for the $c$ electrons).
Furthermore, making a $\pi /2$ rotation around  the
$z$ axis, implemented by
$(\eta^{x},\eta^{y},\eta^{z})\rightarrow(\eta^{y},-\eta^{x},\eta^{z})$, 
we obtain the relations
\begin{equation}\label{RelationsPropagEta}
G_{xy} = - G_{yx}, \qquad G_{xx}=G_{yy}.
\end{equation}
The Hamiltonian is also invariant under a $\pi $ rotation around the $x$ axis 
($(\eta^{x},\eta^{y},\eta^{z})\rightarrow(-\eta^{x},\eta^{y},\eta^{z})$)
together with a change of sign of the magnetic fields
$B_{s}$ and $B_{c}$, and hence
 $G_{xy}$ and $G_{{xx}}$ are respectively odd and even in the
magnetic field. In particular, for $B_{s}=B_{c}=0$, $G_{xy}=0$.
The foregoing arguments apply to the full propagator and to the free
propagator $G^{0}$ (computed with only the magnetic field $B_{s}$).

A different basis is also useful: defining the Dirac fermion $f$ and its
Green's function $G_{B}$ by
\begin{align}
f&\equiv\frac{\eta^{x}- i \eta^{y}}{\sqrt{2}},\\
G_{B} (t) &= -i \dkmoy{f (t) f^{\dagger } (0)},
\end{align}
we have the relations: 
\GroupeEquations{\label{RelGBGXY}
\begin{align}
G_{xx} (\omega ) & = \hphantom{-} G_{yy} (\omega ) = \frac{1}{2}\left(G_{B}
(\omega ) + G_{-B} (\omega ) \right),\\
G_{xy} (\omega ) & = - G_{yx} (\omega ) = \frac{i}{2}\left(G_{B}
(\omega ) - G_{-B} (\omega ) \right).
\end{align}
}
In the $(f,\eta^{z})$ basis, the propagator is diagonal, so it
is more convenient for example to write the Dyson equation, whereas
the original basis $(\eta^{x},\eta^{y},\eta^{z})$ is more convenient
for the diagrammatics.
The bare propagators in the  $(f,\eta^{z})$ basis are given by:
\GroupeEquations{\label{BarePropagatorFreeSpin} 
\begin{align}
R_{z}^{0} (\omega ) &= \frac{1}{\omega +i s},\\
K_{z}^{0} (\omega ) &= 2i h_{\rm eq} (\omega )\frac{s}{\omega^{2} +s^{2}},\\
R_{B}^{0} (\omega ) &= \frac{1}{\omega -B +i s},\\
K_{B}^{0} (\omega ) &= 2i h_{\rm eq} (\omega )\frac{s}{(\omega-B)^{2} +s^{2}},
\end{align}
}
where $s$ is a small regulator which, as discussed in section \ref{PrincipleComputation}
above, should be thought of as the width due to coupling to an auxiliary thermal bath.

\subsubsection{Vertex factors and Dyson equations}\label{sub.vertexANDdyson}

\begin{figure}[hbt]
\[
\fig{5cm}{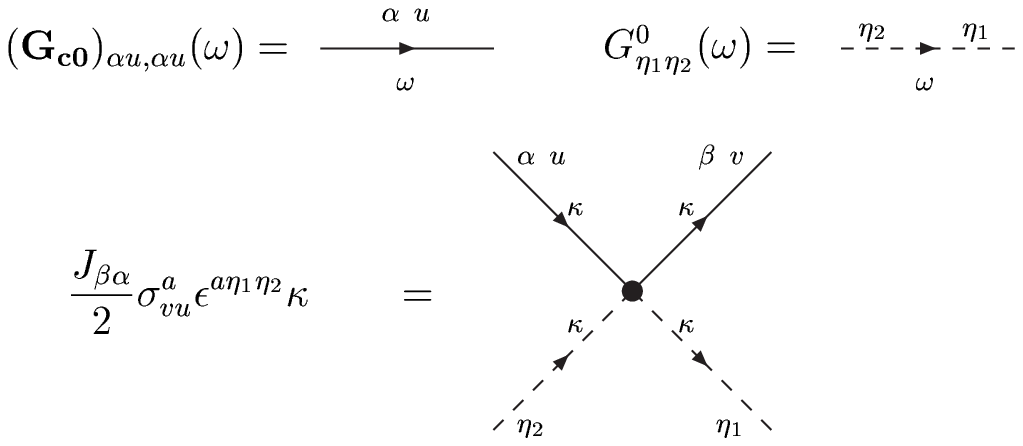}
\]
\caption{\label{Fig.FeynmanRules}
\sl 
Feynman rules: we represent the Majorana fermions with dashed lines, the
electrons with solid lines. $\alpha ,\beta $ are lead indices, $u,v$
are spin indices for the electrons, $\kappa$ is a Keldysh index
$\kappa =\pm$, $\eta_{i},a \in \left\{x,y,z\right\}$ are flavor indices for the
Majorana fermion, and $\sigma$ are the Pauli matrices. The vertex is given in the $\pm$ basis.
}
\end{figure}

The vertex factors can be extracted simply from the
Hamiltonian, and the
Feynman rules are summarized in Fig.~\ref{Fig.FeynmanRules} (in the
$\pm$ basis).  Note that we have oriented the Majorana fermion lines, despite
the operator property that $(\eta^\dagger)\,{=}\,\eta$; a proof that the
lines are orientable is given in Appendix \ref{app.orientability}.

We now derive the expression for the Dyson equation describing the $\eta$ fields.
In general a Dyson equation reads 
\begin{equation}\label{Dyson1}
{\bf G}^{-1} (\omega ) = {\bf G_{0}}^{-1} (\omega ) - \boldsymbol{\Sigma}
(\omega ),
\end{equation}
where the inversion has to be taken in the tensor product of the $x,y,z$
space and the Keldysh space (thus with $6\times 6$ matrices). The free
propagator is given by (\ref{BarePropagatorFreeSpin}).
In the  $x,y,z$ basis, the  self energy can be written as: 
\begin{equation}\label{SelfEta1}
\boldsymbol{\Sigma} (\omega ) = 
\matrice{   \Sigma_{d} (\omega )    &  i  \Sigma_{a} (\omega ) & 0\\
           -i \Sigma_{a}  (\omega ) &  \Sigma_{d} (\omega ) & 0 \\
               0  &  0 &  \Sigma_{z} (\omega )}
\qquad 
\Sigma_{\mu } (\omega ) = 
\matrice{ \Sigma^{R}_{\mu } (\omega ) & \Sigma^{K}_{\mu } (\omega )\\
                0  &  \Sigma^{A}_{\mu }(\omega )} \quad 
\mu \in \{a,d,z \}.
\end{equation}
Using the definitions (\ref{RelGBGXY}), we find:
\begin{equation}\label{DefGB}
G_{\pm B} =\matrice{ 
\dfrac{1}{\omega \mp B - \Sigma_{\pm B}^{R} (\omega ) + is } & 
\dfrac{\Sigma^{K}_{\pm B} (\omega ) +  2i s  h_{\rm eq} (\omega )}
{\left|\omega \mp B  - \Sigma_{\pm B}^{R}  (\omega
) + i s \right|^{2}} \\
\\
0 & \dfrac{1 }{\omega \mp B - \Sigma_{\pm B}^{A}  (\omega ) - is } 
}
\qquad \qquad 
\Sigma_{\pm B} (\omega ) \equiv  \Sigma_{d} (\omega ) \pm \Sigma_{a}
(\omega ).
\end{equation}

\subsection{Perturbation expansion in the steady state}\label{sub.Perturbationsteadystate}
After these preliminaries, let us now come back to the perturbative expansion in the steady
state and develop on the points introduced in
\ref{PrincipleComputation}.  In paragraph \ref{sub.NonCommute} we show that the limits
$J \to 0$ and $s \to 0$ do not commute; in paragraph
\ref{sub.derivationEq4} we solve that problem and present the
derivation of Eq.~(\ref{MAINRESULT}) together with an algorithm for computing higher order
terms in the perturbative expansion.

\subsubsection{Non-commutativity of limits}\label{sub.NonCommute}
We are interested in the steady-state values of the magnetizations
 (\ref{DefMdot}) and (\ref{DefMleads}), which can be expressed in the following
way:
\GroupeEquations{\label{HowToExpressPhysicalQuantities}
\begin{align}
M_{\text{dot}} &= \frac{i}{4} \int \frac{d\omega }{2\pi }
\Bigl ( K_{B} (\omega ) - K_{-B} (\omega )\Bigr ), \label{HowToExpressMdot} \\
M_{\text{leads}} &= \frac{i}{2} \int \frac{d\omega }{2\pi }
\sum_{\empile{\alpha =R,L}{u=+,- }}
(-1)^{u} ({\bf  K_{c}})_{\alpha u,\alpha u},
\end{align}
}
Indeed, the dot magnetization is given by: 
\[
M_{\text{dot}} = \moy{S^{z} } = -\frac{i}{2}
\moy{\left[\eta^{x},\eta^{y} \right]} = 
\frac{1}{2}K_{xy} (t=0).
\]
$M_{\text{leads}}$ is derived analogously.

We begin with the expression for the dot magnetization derived from
(\ref{HowToExpressMdot}) using the Dyson equation (\ref{DefGB}): 
\begin{equation}\label{SuperFormulePourMdot}
M_{\text{dot}} = \frac{1}{4\pi } \int d \omega \,  \Im
\left(\frac{h_{\rm eq} (\omega )}{\omega -B - \Sigma_{B}^{R} (\omega ) + is} \right)
-
\int \frac{d\omega }{8i\pi} \frac{\Sigma_{B}^{K} (\omega ) - h_{\rm eq} (\omega
)[\Sigma^A_{B} (\omega )-\Sigma^R_B(\omega)]}{\left| \omega -B - \Sigma_{B}^{R} (\omega ) +
is\right|^{2}} - (B\rightarrow -B),
\end{equation}
where 
the last term denotes an antisymmetrization in $B$.
The first term in (\ref{SuperFormulePourMdot}) can be transformed
into a sum over Matsubara frequencies and thus we see that
the limits $J\rightarrow 0$ and $s\rightarrow 0$ commute in that term. However, the
second term is more interesting. Firstly,
it vanishes in equilibrium since the numerator cancels, as required by the FDT.
Moreover, using (for $\epsilon \rightarrow 0$)
\begin{equation}\label{DeltaDivergence}
\frac{1}{\left| \omega - B  + i \epsilon \right|^{2} }
\sim \frac{1}{\epsilon } \delta \left( \omega - B  \right),
\end{equation}
we see that in that second term these limits do not commute:
\begin{itemize}
\item $J\rightarrow 0$ and then  $s\rightarrow 0$.  We first take
$\Sigma_{B}^{R}=0$ in the denominator.
We find that this second term in (\ref{SuperFormulePourMdot}) gives a diverging term at second order in $J$
proportional to $J_{RL}^{2}/s$. More precisely, using the expression for the self-energy at second
order derived in Appendix \ref{App.Sigma}, and (\ref{DeltaDivergence})
with $\epsilon =s$, we find
\[
M_{\text{dot}} \sim  \frac{J_{RL}^{2}}{s} \tanh \left(\frac{B}{2T} \right)
\Biggl[
\varphi \left(\frac{B}{T} \right)  -
\frac{1}{2}
\Biggl (
\varphi \left(\frac{B+V}{T} \right)  + \varphi \left(\frac{B-V}{T} \right)
\Biggl)
\Biggr]  
\]
up to a finite (i.e.\ not diverging as
$s \to 0$) term of order $J^{2}$ and to $O (J^{3})$ terms ($\varphi$
is defined in (\ref{DefinitionPhi})).
\item $s\rightarrow 0$ and then $J\rightarrow 0$. Using
(\ref{DeltaDivergence}) with $\epsilon =\Im \,\Sigma^R_B (B)$ (we work at dominant order in $J$), we find: 
\begin{equation} \label{trueMdot}
M_{\text{dot}} = - \frac{1}{4} \left( \frac{\Sigma^{K}_B (B)}{\Sigma^A_B (B)
- \Sigma^R_B(B)} \right) -
(B\rightarrow -B) + O (J)
\end{equation}
where the self-energy has to be expanded at order 2 in $J$ (the first
term in (\ref{SuperFormulePourMdot}) cancels a part of the second term).
\end{itemize}
The second limit is the physical one (as explained in section
\ref{s:model})
and it gives a formula (\ref{trueMdot}) for the magnetization at order 0.
However, to make use of this formula one needs to know the Keldysh ($\Sigma^K_B(B)$) and
spectral ($\Sigma^A_B(B)-\Sigma^R_B(B)$) parts of the self-energy.  The leading terms of
these self-energies are of order $2$ in $J$, but because the ratio of them is taken, they
determine the magnetization at order $0$.  The crucial point is that (\ref{trueMdot}) is
in fact an implicit equation for $h_B(B)$, which appears on the left-hand side because
$M_{\rm dot}$ is defined in terms of it by (\ref{HowToExpressPhysicalQuantities}),
and on the right-hand side because the self-energies at $O(J^2)$ depend on the Keldysh
Green's function at $O(1)$, and hence on $h_B$.
The most elegant way to capture this `feedback' effect is to reformulate the problem
in terms of the perturbative expansion of the steady-state Green's function in powers of
$J$, starting from the Dyson equation in which the $s \to 0$ limit has already been taken.
We now describe this method in more detail, and specify the procedure for calculating
the steady-state Green's functions perturbatively in $J$ to arbitrary order.

\subsubsection{Perturbative expansion of steady-state quantities}\label{sub.derivationEq4}

Let us now give a general method for computing the perturbative expansion
of the Green's functions, and use it to derive Eq.~(\ref{MAINRESULT}).
First, we reformulate slightly the diagrammatic expansion in
terms of the full Green's function ${G_{\psi}}$ and of skeleton
diagrams, and we explicitly derive the steady-state Dyson equation as
a functional equation for $G_{\psi}$ (again, for simplicity, we write
some equations for a generic field $\psi$). The ``generic'' Dyson equation reads: 
\begin{equation}\label{DysonEquation}
{G_{\psi }} (J,s,\omega ) = 
\left(
G_{\psi 0}^{-1} (\omega ) - \Sigma_{ \text{skel}}[G_{\psi }] (J,s,\omega )
 \right)^{-1},
\end{equation}
where $\Sigma_{\text{skel}}$ is a functional of $G_{\psi}$
defined by the skeleton  expansion of the self-energy (see Appendix \ref{App.DerivationDyson} for a derivation).
We first take the $s\rightarrow 0$ limit in
(\ref{DysonEquation}) and then solve it order by order in $J$.
It is in principle sufficient to solve (\ref{DysonEquation})
for $G_{\psi}$ since the magnetization can be directly extracted from a
Green's function, and other physical quantities (e.g. currents) are given by their skeleton expansions.

We first derive the explicit form of the Dyson equation for all
fields. Defining
\begin{equation}\label{DefPetitSigma}
\sigma_{\psi }^{K} \equiv \frac{\Sigma^{K}_{\psi }}{2i\pi },
\qquad 
\sigma_{\psi } \equiv \frac{\Sigma^{A}_{\psi } - \Sigma^{R}_{\psi }}{2i\pi },
\end{equation}
the full set of  Dyson equations for the electrons and the Majorana fields
can be rewritten explicitly as:
\begin{align}
{\bf R_{c}} (\omega ) &= \frac{1}{{\bf R}^{-1}_{\bf c0} - \boldsymbol{\Sigma_{R}}
(\omega )},\label{DysonElectronR}\\ 
{\bf K_{c}}(\omega) &= \left( 1 - {\bf R_{c0}} \boldsymbol{\Sigma_{R}} \right)^{-1}
{\bf K_{c0}} \left( 1 - \boldsymbol{\Sigma_{A}} {\bf A_{c0}} \right)^{-1}
+ \left( {\bf R}^{-1}_{\bf c0} - \boldsymbol{\Sigma_{R}} \right)^{-1} \boldsymbol{\Sigma_{K}}
\left( {\bf A}^{-1}_{\bf c0} - \boldsymbol{\Sigma_{A}} \right)^{-1}
\label{DysonElectronK}\\
R_{B} (\omega ) &= \frac{1}{\omega - B - \Sigma_{B}^{R} (\omega ) +i0^{+}},
\label{DysonSpinR}\\
K_{B} (\omega ) &= h_{B} (\omega )\Bigl ( A_{B} (\omega ) - R_{B}
(\omega )   \Bigr),
\label{DysonSpinK}\\
h_{B} (\omega ) &\equiv  \frac{\sigma_{B}^{K} (\omega )}{\sigma_{B}
(\omega )},
\label{DefhB}\\
R_{z} (\omega ) &= \frac{1}{\omega  - \Sigma_{z}^{R} (\omega ) +i0^{+}},
\label{DysonSpinzR}\\
K_{z} (\omega ) &= h_{z} (\omega )\Bigl ( A_{z} (\omega ) - R_{z}
(\omega )   \Bigr),
\label{DysonSpinzK}\\
h_{z} (\omega ) &\equiv  \frac{\sigma_{z}^{K} (\omega )}{\sigma_{z}
(\omega )},
\label{Defhz}
\end{align}
where the bold symbols are $4\times 4$ matrices (in lead-spin
space). For completeness, we have also written the definitions of the $h$
functions. We have three blocks of equations, for the
electrons, the $f$ field and the $\eta^{z}$ field respectively. Within these
blocks, we have an equation for the retarded function
(\ref{DysonElectronR},\ref{DysonSpinR},\ref{DysonSpinzR}), an
equation for the Keldysh function
(\ref{DysonElectronK},\ref{DysonSpinK},\ref{DysonSpinzK}),
 and for the Majorana fermions the definition of the $h$ function
(\ref{DefhB},\ref{Defhz}).
We  define $h_{B}$ with (\ref{DefhB}) rather than with (\ref{DysonSpinK}) since
the spectral density is a delta peak at order 0, whereas the self-energy
is a smooth function.

\begin{figure}[ht]
\[
\figx{5cm}{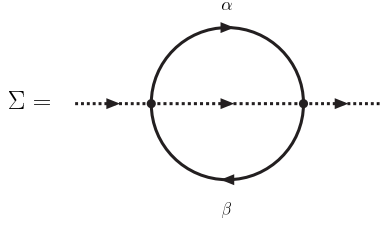}
\]
\caption{\label{Fig.DiagSigma}
\sl 
Diagram of the skeleton self-energy at second order.
}
\end{figure}

The spin and the lead electrons appear on a different footing: the order 0 part of the electronic Keldysh function 
is given by ${\bf K_{c0}}$ whereas the order 0 parts of $ h_{B}$ and $ h_{z}$ must be computed using equations 
(\ref{DefhB},\ref{Defhz}). Note that to obtain the order $n$ part of $h_{B}$, one needs to compute the
self-energies at order $n+2$.
The method of obtaining the perturbative expansion in $J$ order by
order is as follows (denoting by $f^{(n)}$ the order $n$ part of any function $f$):
\begin{enumerate}
\item Assume that we have the expansion of all functions to
order $n-1$.
\item Since $\Sigma_{\text{skel}}$ is at least of order 1, using
(\ref{DysonElectronR},\ref{DysonSpinR},\ref{DysonSpinzR}), we
compute  $R_{B}^{(n)}$, $R_{z}^{(n)}$,  ${\bf R_{c}}^{(n)}$ and ${\bf K_{c}}^{(n)}$.
\item Since $\Sigma_{\text{skel}}^{K}$ and $\Sigma_{\text{skel}}''$ begin at order 2, we compute
the $\sigma^{K}$ and $\sigma$ to order $n+2$, as functions of
the unknowns $h_{B}^{(n)}$ and $h_{z}^{(n)}$.
\item We then obtain closed equations for  $h_{B}^{(n)}$ and $h_{z}^{(n)}$ from 
(\ref{DefhB},\ref{Defhz}) (expanded to order $n$).
\end{enumerate}
Thus the order 0 part of the impurity magnetization  is
given by (from (\ref{HowToExpressMdot})): 
\begin{equation}\label{MdotAsAFunctionofHb}
M_{\rm dot} = - \frac{1}{2} h_{B}^{(0)} (B).
\end{equation}
The order 0 parts of $R_{B}$ and $K_{B}$ are
\begin{equation}
R^{(0)}_{B}=\frac{1}{\omega -B + i0^{ +}}, \qquad 
K^{(0)}_{B} = 2i\pi h^{(0)}_{B} (B) \delta (\omega -B),
\end{equation} 
and the bare Green's functions of the electrons are given by
(\ref{GelectronOrder0}).
We compute  $h_{B} (\omega= B )$ at order 0  by
expanding the self-energies at second order (given by the diagram
of Fig.~\ref{Fig.DiagSigma}), and then solving (\ref{DefhB})
for $h_{B} (B)$. Finally we find (\ref{MAINRESULT}) in the large bandwidth limit ($D\rightarrow \infty$).
The computation is presented in detail in Appendix \ref{App.Sigma}.
This completes the computation of the Green's function to order 0.

Note that ${\bf K_{c}}$ is thermal at order 0, which expresses
the fact that the leads are in thermal equilibrium.  Therefore the leads'
contribution to the total magnetization at order 0 is given by the
Pauli term, which was explicitly excluded from $M_{\text{tot}}$;
hence $M_{\text{tot}}= M_{\text{dot}}$ at this order, as claimed in
(\ref{MAINRESULT}). Moreover, we have not written explicitly the full forms of the functions
$h_B(\omega)$ and $h_z(\omega)$ to this order, since they are not required
in the calculation of the zeroth-order Keldysh functions:\ $h_B(B)$ and $h_z(0)=0$
are sufficient, since the spectral density (at this order) is a delta peak.
The full functions would, however, be needed to compute at second order;
the function $h_B(\omega)$ can easily be extracted from Appendix \ref{App.Sigma},
and the calculation of $h_z(\omega)$ proceeds along similar lines.

Our earlier interpretation of the $1/s$ divergences in
`straightforward' perturbation theory
is borne out by this result.  As stated above, the $1/s$ divergences result from
an incorrect choice of zeroth-order distribution function $h$.  If we insert
the $O(1)$ part of (\ref{DefhB}) into (\ref{SuperFormulePourMdot}) in place
of $h_{\rm eq}$, we see that the divergences are cancelled at order $J^2$, since
to this order we are now using the correct long-time distribution function.  This
shows that our method and that of \cite{WingreenSivan} are equivalent.

\begin{figure}[ht]
\[
\figx{13cm}{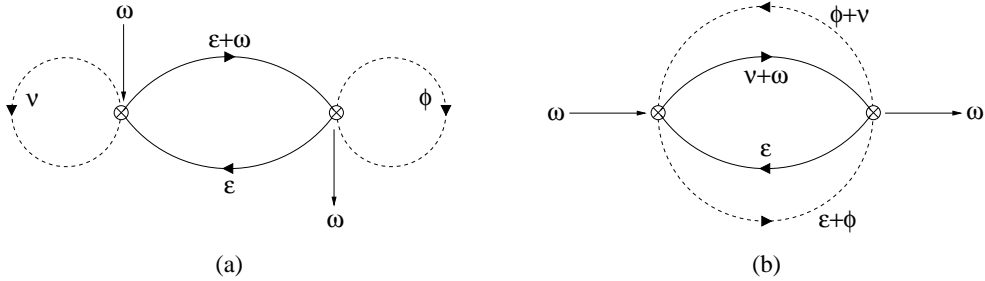}
\]
\caption{\label{Fig.cccorrel}
\sl 
The two skeleton diagrams contributing to the current-current correlator
at leading order.  In each diagram, one of the electrons (represented by the
solid lines) is from the left lead, and the other is from the right.
The vertices represent $I$, the current operator.
}
\end{figure}

It is important to note that these corrections to the zeroth-order terms in
perturbation series are in no way restricted to quantities such as the
magnetization.  On the contrary, since what we have really calculated is the
correction to the zeroth-order Keldysh Green's function, they manifest themselves
in many quantities.  As an example, we may consider the current-current correlator:\ the
leading terms in this quantity are of order $J^2$, and
are calculated by inserting the zeroth-order Green's functions into
the skeleton diagrams shown schematically in Fig.~\ref{Fig.cccorrel}.
The result is that
\begin{eqnarray}
\left\langle I(-\omega) I(\omega) \right\rangle
& = & \frac{\pi \left( \rho_0 J_{LR} \right)^2}{2 \hbar^2}
\Bigg\{ \left[ h^{(0)}_B(B) \right]^2 p(\omega,V)
- h^{(0)}_B(B) \left[ p(\omega+B,V) - p(\omega-B,V) \right]
\nonumber \\
& \ & \qquad \qquad \qquad
+ \left[ p(\omega,V) + p(\omega-B,V) + p(\omega+B,V) \right]
\Bigg\}, \label{cccorrel}
\end{eqnarray}
where the function $p(x,y)$ is defined by
\begin{equation}
p(x,y) = \frac{x-y}{e^{\beta(x-y)}-1}
+ \frac{x+y}{e^{\beta(x+y)}-1}.
\end{equation}
The result (\ref{cccorrel}) is a function of
$h^{(0)}_B(B)$, and is therefore clearly sensitive
to the corrections made to the zeroth-order Green's function, as
expected on the basis of the discussion above.

\section{Conclusion}

The purpose of this note has been to present the expression
for the order 0 magnetization in the Kondo model out of equilibrium
(Eq.~\ref{MAINRESULT}), and a systematic procedure for obtaining
higher-order corrections to this result. 
The result we obtain may seem surprising, in the sense that even at order 0 in $J$ it
does not coincide with the equilibrium expression $ \frac{1}{2} \tanh
(B/2T)$. Indeed the out-of-equilibrium distribution function which describes
the population of the two levels of a weakly coupled spin is in
general not thermal but must be computed by solving a transport
equation: it is determined by the steady state into which 
the voltage difference forces the system. 
Moreover, this distribution function also enters the computation of other physical quantities (e.g. currents)
and their perturbative expansions therefore exhibit similar phenomena.

Finally, we stress that the issue discussed in this note is not directly
related to the so-called ``decoherence time'' issue.  Answering the
question of strong coupling at finite voltage requires computations at
higher orders in $J$.  It is clear from the above, however, that the
behavior of the out-of-equilibrium perturbative expansion will be
markedly different from that of its equilibrium counterpart.  For example,
the $O(J)$ contribution to the Keldysh Green's function of the Majorana
fermions depends on the $O(J^3)$ contribution to the self-energies.
These, however, contain terms that diverge like $\ln D$ in the $D \to
\infty$ limit, and so such logarithmic divergences may be expected to
appear at $O(J)$ in some of the Green's functions, and therefore in
physical properties such as the magnetization.  Indeed, this phenomenon
has been reported recently \cite{AchimPreprint}; the full
interpretation of this striking departure from equilibrium behavior merits
further work.

\acknowledgments
This work has been supported by the Center for Materials Theory at Rutgers University,
by the EPSRC (UK) under grant number GR/M70476,
and by the U.S.\ Department of Energy under grant number DOE grant DE- FG02 - 00ER45790.
We thank L. I. Glazman and A. Kaminski for useful discussions and 
for the master equation derivation presented in Appendix
\ref{RelBoltzmann}.
We also thank E. Abrahams, N. Andrei, G. Biroli, R. Chitra,
G. Kotliar, D. Langreth, Y. Meir, B. Mouzykantskii, A. Ruckenstein and A. Schofield
for useful discussions. 
Our thanks are also due to the first author of ref.~\cite{WRONGPRL}, P. Coleman,
for numerous discussions, though we note that he is in disagreement with the corrected analysis 
presented above \cite{ColemanWrong},
and continues to support the perturbative result presented in Eq.~(2) of
ref.~\cite{WRONGPRL}.
During the completion of this paper, we learned that the same conclusion
(Eq. 4) has been independently obtained by A. Rosch, P. W{\"o}lfle and J. Kroha.
Part of this work was completed during our stay at the Aspen Center for Physics in July 2001.


\appendix

\section{A semi-classical derivation of (\ref{MAINRESULT})}\label{RelBoltzmann}
The magnetization at order 0 in $J$ (\ref{MAINRESULT}) can also be derived
using a semiclassical master equation approach \cite{GlazmanKaminskyPrivate}.
By symmetry, the reduced density matrix of the spin is diagonal, so we
consider the spin as classical and characterize its state by the
probability of its being up, $P_{\uparrow}$, or down, $P_{\downarrow}$. The spin
dynamics is governed by a master equation:  
\GroupeEquations{\label{MasterEquation}
\begin{align}
\frac{dP_{\uparrow} (t)}{dt} & = \Gamma_{\downarrow \uparrow}
P_{\downarrow } (t) - \Gamma_{\uparrow \downarrow} P_{\uparrow } (t),\\
\frac{dP_{\downarrow} (t)}{dt} & = \Gamma_{\uparrow \downarrow}
P_{\uparrow } (t) - \Gamma_{\downarrow \uparrow } P_{\downarrow } (t),
\end{align}}
where $\Gamma_{AB}$ is the rate of the spin-flip process
$A \rightarrow B$ induced by the Kondo terms. At
second order in perturbation theory, these rates are given by:
\GroupeEquations{\label{TauSpinflip}
\begin{align}
\Gamma_{\downarrow \uparrow} &\approx 
\sum_{\alpha ,\beta \in \{R,L \}}
J_{\alpha \beta }^{2}
\int d \epsilon_{1} d \epsilon_{2} \;
\rho (\epsilon_{1} - B_{c} - \mu_{\alpha})\rho (\epsilon_{2} + B_{c}- \mu_{\beta })
n_{F} (\epsilon_{1} - \mu_{\alpha})
\Bigl ( 1-  n_{F} (\epsilon_{2} - \mu_{\beta })\Bigr )
\delta \bigl ( \epsilon_{1} - \epsilon_{2} + B_{s}  \bigr )
\\
\Gamma_{ \uparrow\downarrow } &\approx 
\sum_{\alpha ,\beta \in \{R,L \}}
J_{\alpha \beta }^{2}
\int d \epsilon_{1} d \epsilon_{2} \;
\rho (\epsilon_{1} + B_{c}- \mu_{\alpha})\rho (\epsilon_{2} - B_{c}- \mu_{\beta })
n_{F} (\epsilon_{1} - \mu_{\alpha})
\Bigl ( 1-  n_{F} (\epsilon_{2} - \mu_{\beta })\Bigr )
\delta \bigl ( \epsilon_{1} - \epsilon_{2} - B_{s}  \bigr )
\end{align}
}
Taking the large bandwidth ($D \to \infty$) limit and integrating, the rates become:
\GroupeEquations{
\begin{align}
\Gamma_{\uparrow \downarrow } & = 
\Bigl ( J_{R}^{2} + J_{L}^{2}
\Bigr)
\frac{B_{s}}{e^{B_{s}/T} - 1} 
+
J_{RL}^{2}
\left(
\frac{B_{s}-V}{e^{(B_{s}-V)/T} - 1} 
+
\frac{B_{s}+V}{e^{(B_{s}+V)/T} - 1} 
 \right), \\
\Gamma_{\downarrow \uparrow} & = 
\Bigl ( J_{R}^{2} + J_{L}^{2}
\Bigr)
\frac{ -B_{s}}{e^{-B_{s}/T} - 1} 
+
J_{RL}^{2}
\left(
\frac{-B_{s}-V}{e^{-(B_{s}+V)/T} - 1} 
+
\frac{-B_{s}+V}{e^{-(B_{s}-V)/T} - 1} 
 \right).
\end{align}}
The steady state value of the probability is given by equating the
left hand sides of (\ref{MasterEquation}) to 0, and the magnetization
is given by $M_{\text{dot}} = (P_{\uparrow} -P_{\downarrow })/2$, thus 
\begin{equation}\label{RelMdotProba}
\Mdot = \frac{1}{2} \left( \frac{\Gamma_{\downarrow \uparrow}
-\Gamma_{\uparrow \downarrow } }
{\Gamma_{\downarrow \uparrow} + \Gamma_{\uparrow \downarrow }} \right),
\end{equation}
which leads to Eq. (\ref{MAINRESULT}).

One can see that the Dyson equation in steady state, at second order
in $J$, maps exactly to the master equation, although they appear to
have different transients. So at order 0, this computation is just a
reformulation of the one presented above.

\section{A free model}\label{App.Freemodels}
In this appendix, we recall \cite{NozieresCaroli,pertU2} the solution of a simple free model
in an out-of-equilibrium steady state regime, which displays a result
very similar to (\ref{MAINRESULT}).
We consider a free level coupled to some reservoirs of free electrons
via hopping terms.
The Hamiltonian is given by: 
\begin{equation}\label{HamFreeDot}
H  =  \sum_{\alpha {\bf k}} \varepsilon_{{\bf k}} c^\dag_{\alpha
{\bf k}} c_{\alpha {\bf k}} +
\sum_{\alpha} g_{\alpha } \bigl( c^{\dagger}_{\alpha } d + d^{\dagger} 
c_{\alpha } \bigr)
+ 
\epsilon_{d} d^{\dagger }d,
\end{equation}
where the local reservoir-electron operator $c_\alpha$ is defined by
$c_\alpha = c_\alpha({\bf x}={\bf 0})
\equiv \Omega^{-1/2} \sum_{\bf k} c_{\alpha {\bf k}}$.  ($\Omega$ is
the usual normalization factor related to the volume of the reservoirs.)
As before, we assume that the reservoirs are thermalized with the same temperature as each other,
but with different chemical potentials
$\mu_{\alpha}$.
In particular, we are interested in a model with three reservoirs:\ 1 and 2 are the leads
(at different potentials)
and 3 an additional thermal bath (with $\mu_{3}=0$) to which the level is coupled.
Since it is a Gaussian model, we can simply solve the Dyson equation: 
\begin{equation}\label{App.DysonModelLibre}
G^{-1} = G_{0}^{-1} - \Sigma.
\end{equation}
We use the $(d,c_{1},c_{2},c_{3})$ basis. The inverse bare Green's function is given by: 
\begin{equation}
G^{-1}_{0} (\omega ) = 
\begin{pmatrix}
 \omega - \epsilon_{d}  & 0  & 0 &0\\
0   &  G_{1}^{-1}    & 0 & 0 \\
0   & 0 & G_{2}^{-1}  &0 \\
0   & 0 & 0 & G_{3}^{-1}
\end{pmatrix},
\end{equation} 
where $G_{\alpha }$ is the bare Green's function of the reservoir $\alpha$, given in the large bandwidth
limit by (in the LO basis):
\begin{equation}\label{App.GreenFunctionBareLeads}
G_{\alpha}^{ -1} (\omega ) =  \frac{1}{\pi \rho_{0}}
\begin{pmatrix}
 i    &   -2i h_{\rm eq} (\omega  - \mu_{\alpha })\\
0& -i
\end{pmatrix},
\end{equation}
where $\rho_{0}$ is the density of states, which we take to be the same for each reservoir.
The self-energy is given by: 
\begin{equation}\label{App.SelfEnergyModelLibre}
\Sigma  (\omega ) = 
\begin{pmatrix}
0  &  g_{1}  & g_{2}  &g_{3} \\
g_{1}   &  0    & 0 & 0 \\
g_{2}   & 0 & 0  &0 \\
g_{3}   & 0 & 0 & 0
\end{pmatrix}.
\end{equation}
Solving (\ref{App.DysonModelLibre}), we find the occupation of the dot in the steady state:
\begin{equation}\label{App.DEfOccDOT}
n (\epsilon_{d}) \equiv \moy{ d^{\dagger } d}.
\end{equation}
In  the limit where $g_{1}, g_{2}, g_{3}$ go to zero in fixed ratios, we obtain: 
\begin{equation}\label{App.OccupationDotLibre}
n (\epsilon_{d}) = \dfrac{
g_{1}^{2} n_{F} (\epsilon_{d} - \mu_{1}) +
g_{2}^{2} n_{F} (\epsilon_{d} - \mu_{2}) +
g_{3}^{2} n_{F} (\epsilon_{d} )
}{g_{1}^{2} +  g_{2}^{2} + g_{3}^{2}}
\end{equation}
where $n_{F}$ is the Fermi function.
The properties of this result
are similar to those of (\ref{MAINRESULT}):\ the occupation of the dot, in the limit of zero couplings,
depends on the ratios of these couplings and
is not in general given by the Fermi function.
Moreover, if $g_{3}\gg g_{1},g_{2}$, we recover the equilibrium result since the physics is dominated 
by the thermal bath 3. If, on the other hand, we take $g_{3} \ll g_{1},g_{2}$, we find a non-equilibrium result
since the occupation at order 0 in perturbation theory in the $g$'s is determined by the leads 1 and 2.

\section{Orientability of Majorana fermion lines}\label{app.orientability}

In this appendix, we demonstrate that we can treat the Majorana lines as oriented in
the diagrams.
To show this, it is simplest to take the lines to represent the bare
zero-field Majorana functions, and to treat both the magnetic field
and the Kondo interaction as vertices.  We have three species of
Majorana fermion, $\left\{\eta^x,\eta^y,\eta^z\right\}$; formally, we
may represent each of these as the sum of the creation and annihilation
operators of a Dirac fermion:
\begin{equation}
\eta^a = \frac{1}{\sqrt{2}}  \left( f_a + f^\dagger_a \right);
\end{equation}
the dual operator $f_a - f^\dagger_a$ decouples from the problem and
may be ignored.
In the absence of a magnetic field and interactions, all Majorana
Green's functions are diagonal, so the only Green's functions that
occur are:
\begin{eqnarray}
G_a & = & -i \dkmoy{\eta^a(t) \eta^a(0)} \nonumber \\
& = & - \frac{i}{2} \dkmoy{f_a(t) f^\dagger_a(0) + f^\dagger_a(t) f_a(0)} \nonumber \\
& = & \left( - \frac{i}{2} \dkmoy{f_a(t) f^\dagger_a(0)} \right) + \mbox{p.h.}, \label{parthole}
\end{eqnarray}
where the notation ``p.h.'' stands for ``particle-hole'', i.e.\ $f \leftrightarrow
f^\dagger$.
We thus see that the bare Majorana Green's function may be written simply as
the sum of two bare Dirac fermion Green's functions of opposite orientations.

It is easy to see that the vertex factors at the magnetic field and Kondo vertices
do not depend on whether the Dirac fermion lines are incoming or outgoing.
In the case of the magnetic field, for example, the interaction term is
\begin{eqnarray}
H_{\rm mag} & = & -B S^z \nonumber \\
& = & iB \eta^x \eta^y \nonumber \\
& = & \frac{iB}{2} \left( f_x + f^\dagger_x \right) \left( f_y + f^\dagger_y \right),
\end{eqnarray}
so we see that the vertex factor is the same irrespective of the orientation
of the two $f$-lines; a parallel argument may be given for the Kondo interaction
vertex.  Hence, each diagram consists of a sum of $2^N$ diagrams that
differ only in the orientations of their $N$ Dirac fermion lines.

But these orientational differences do not alter the value of the diagram, since
the bare Green's function of the $f$-fermion is particle-hole symmetric, and
hence (\ref{parthole}) corresponds to the Green's function for a single orientation
of the Dirac fermion line, with the prefactor $1/2$ removed.  Hence we
may represent the Majorana Green's functions in all diagrams using oriented lines.

\section{Erratum to \cite{WRONGPRL}}\label{App.pointsplitting}

In a previous paper \cite{WRONGPRL}, a formula (Eq. 2) was proposed for the (second order) perturbative expansion
of the magnetic susceptibility in the out-of-equilibrium steady state which was of the form:
\begin{equation}
\chi (T,V) = \frac{1}{4T} +  O (J).
\end{equation}
In this appendix, we briefly rediscuss its derivation  
and explain why it is incorrect.
The method used was straightforward Keldysh perturbation theory
to second order with $s$ finite in which the bare Keldysh function of
the Majorana spin was taken to be thermal. The $1/s$ divergences were regulated
using a ``point splitting'' procedure: since they occur due to the
coincidence of two poles in the integrals, one splits these poles on
the real axis to a distance $\delta$ and sends $\delta$ to 0 at the
end of the computation, term by term. The justification given in
footnote 20 of \cite{WRONGPRL} is however incorrect: endowing the Majorana fermions on the dot
with a fictitious dispersion does not lead to this prescription, since divergences reappear when the
bandwidth is sent to zero.
The ``point splitting'' regularization prescription can however be
described physically as follows: let the applied magnetic fields $B_s$ and $B_c$ oscillate slowly at a frequency
$\omega_0$; compute the total magnetization at frequency $\omega_0$ up to second order
in $J$; then take the limit $\omega_0 \to 0$ in the coefficient of each power of $J$.

However, this regulation procedure is based on an interchange of the order of the
limits $\omega_0 \to 0$ and $J \to 0$.  We want to calculate the static magnetic response,
and consequently wish to take $\omega_0 \to 0$ before $J \to 0$; but in fact the technique
used in \cite{WRONGPRL} does the opposite:\ it expands in $J$ ($J \to 0$) {\em before\/}
taking the $\omega_0 \to 0$ limit.  It is simple to show that these limits do not
commute;
this can be seen explicitly from the form of the second term of
(\ref{SuperFormulePourMdot}):
\begin{equation}\label{App.toyintegral}
\mbox{Re} \, \int d\omega  \frac{g (\omega )}{\bigl ( \omega  + \omega_{0}  + i
\gamma \bigr )\bigl ( \omega  - \omega_{0}  - i \gamma \bigr )},
\end{equation}
where $g$ is some function and $\gamma \sim \Sigma'' (B)$ (the limit $s\rightarrow 0$
having already been taken).
If we take the limits $J\rightarrow 0$ (i.e.\ $\gamma \to 0$) and then $\omega_{0}\rightarrow 0$ we obtain
the result of the point splitting prescription of \cite{WRONGPRL}:
\begin{equation}
\int d \omega  \, \left( \frac{g' (\omega )}{\omega}  \right),
\end{equation}
whereas if we take them in the opposite order, $\omega_{0}\rightarrow 0$ and then 
$J\rightarrow 0$, we recover (\ref{MAINRESULT}).

\section{The steady state Dyson equation}\label{App.DerivationDyson}

In this appendix, we present a formal derivation of
the steady state Dyson equation (\ref{DysonEquation}). 
First of all, let us emphasise that our calculation is carried out
{\em in the steady state regime\/}, that is in the long-time limit after
the switching on of the interactions.  We assume that, in this limit,
the system evolves to a time-independent steady state
under the time evolution described by its Hamiltonian (as required by
the Keldysh method: see section \ref{KeldyshMethodGal}).
Strictly speaking, this is not possible, since the Hamiltonian (\ref{ham}) conserves the total magnetization
of the system and that conservation law prevents the magnetization of
the system from relaxing. However, this conservation law is not physical (we have omitted, for example,
spin-orbit terms in the leads); therefore, to allow the system to relax to its steady state, we 
introduce a coupling $g$ that breaks the conservation laws. 
As a specific example, one could consider an  anisotropic ($J_x \ne J_y \ne J_z$) Kondo model.
As this extra coupling is relaxed to zero, the transient time taken to reach the steady state
diverges but we make the  assumption that the values of physical quantities in the steady
state are smooth functions of $g$.  Therefore, once we have
taken the $t \to \infty$ limit, we can set $g=0$ in 
the equation which determines the steady state.

The details of the derivation are as follows:
\begin{enumerate}
\item Using the perturbative expansion, we establish the Dyson
equation at finite times:
\begin{equation}\label{DysonEqFiniteTimes}
 \int du \,\, 
\Bigl(
G_{\psi 0}^{-1} (t,u) - \Sigma_{ \text {skel}}[G_{\psi } ] (t,u)
\Bigr)
  * G_{\psi } (u,t') = \delta (t-t')\otimes 1,
\end{equation}
where $\Sigma_{\text{skel}}$ is a functional of $G_{\psi}$
defined by the skeleton  expansion of the self-energy. The
product should be understood as a matrix product in the LO basis. 
$G_{\psi } $ is a function of two times and of $J,g$.
To obtain this equation, we write the Dyson equation in the
finite time diagrammatic expansion outlined in section \ref{sub.vertexANDdyson}
\cite{RammerSmith}, and use the definition
of the skeleton diagrams \cite{NozieresPBNCORPS,AGD}.
In (\ref{DysonEqFiniteTimes}), the times $t,t'$ and  $u$ run from $-\infty $
to $\infty$, and the couplings $J$ are time dependent: $J_{\alpha}
(t) = J_{\alpha }\theta (t)$, i.e.\ we switch on the interaction
suddenly at $t=0$.
\item The assumption that the system relaxes to a
non-equilibrium time translation invariant steady state (the coupling
$g$ to the relaxation bath is finite) is transcribed mathematically as
the existence of the limit
\begin{equation}\label{LimiteSS}
\lim_{\empile{t,t'\rightarrow \infty}{t-t'=\tau}} {G (t,t')} = G (\tau )
\end{equation}
(we shall denote it with the same function name).
This assumption is not trivial. In particular,
in \cite{ColemanHooleyLargeNOscillations}, the existence
of steady oscillating states has been suggested using a large-$N$ slave boson treatment.
We exclude them here on physical grounds, since we do not expect such
states to appear in the regime where the perturbation theory is
applicable anyway (i.e.\ at high temperatures or at high magnetic
fields in the Kondo problem). 
We can thus take the long time limit, and Fourier transform the Green's functions and
self-energy in
(\ref{DysonEqFiniteTimes}) to obtain: 
\begin{equation}\label{Dyson2}
{G_{\psi }} (J,s,g,\omega ) =
\left(
G_{\psi 0}^{-1} (\omega ) - \Sigma_{ \text{skel}}[G_{\psi }] (J,s,g,\omega )
 \right)^{-1}.
\end{equation}
\item Taking the $g\rightarrow 0$ limit as discussed above, 
we finally obtain the Dyson equation (\ref{DysonEquation}): 
\begin{equation}
{G_{\psi }} (J,s,\omega ) = 
\left(
G_{\psi 0}^{-1} (\omega ) - \Sigma_{ \text{skel}}[G_{\psi }] (J,s,\omega )
 \right)^{-1}.
\end{equation}
\end{enumerate}

\section{Computation of the self-energy diagram}\label{App.Sigma}

In this appendix, we present the computation of $h_{B} (B)$ at order 0.
The computation is in three steps.  First, we compute
the self-energy diagrams at second order as a function of a ``generic'' diagram.
(This simplifies the problem by separating the Majorana and spin indices
from the Keldysh structure.)
Second, we compute this generic diagram. Finally, we solve the resulting implicit
equation for $h_{B} (B)$. Throughout this appendix,
we streamline our notation by omitting the temperature $T$; in the final formulas,
therefore, $B$ and $V$ should be replaced by $B/T$ and $V/T$ respectively.

\begin{figure}[ht]
\[
\figx{8cm}{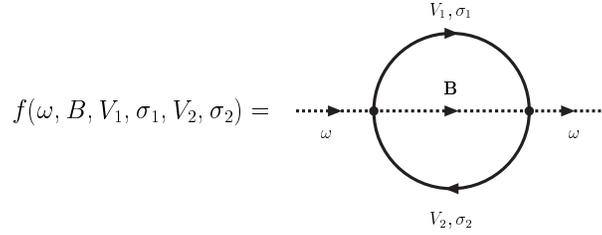}
\]
\caption{\label{Fig.DiagSigmaGeneric}
\sl 
Diagram of the ``generic'' self-energy at second order.
}
\end{figure}

\begin{figure}[ht]
\[
\figx{10cm}{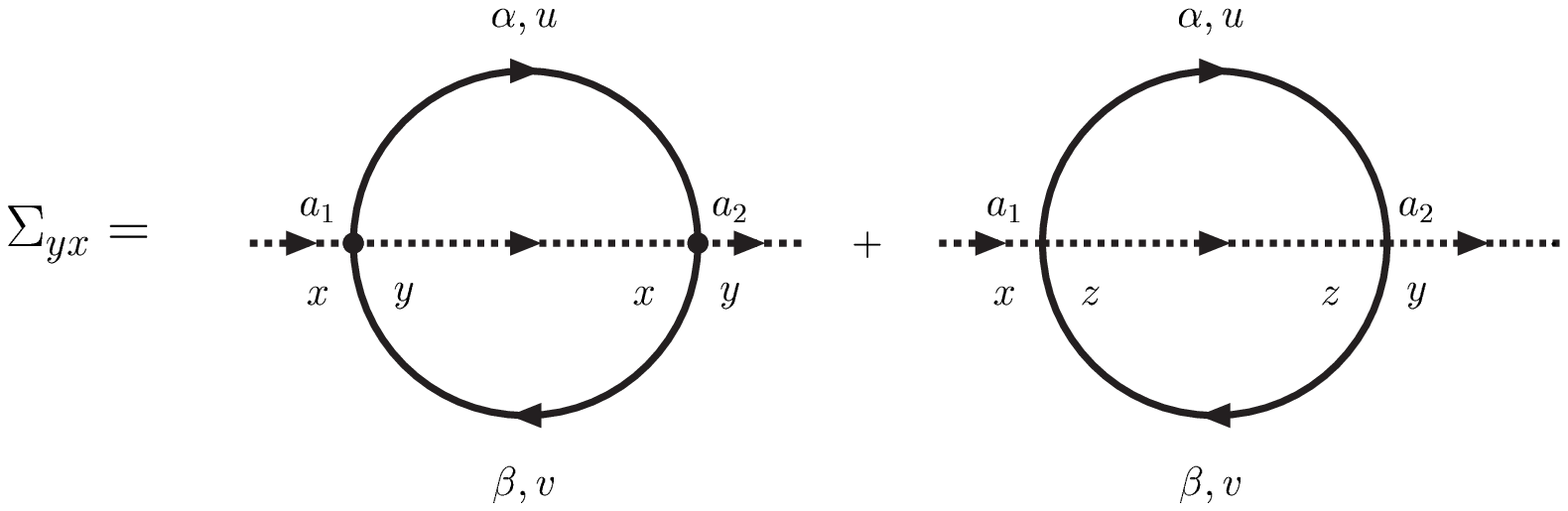}
\]
\[
\figx{10cm}{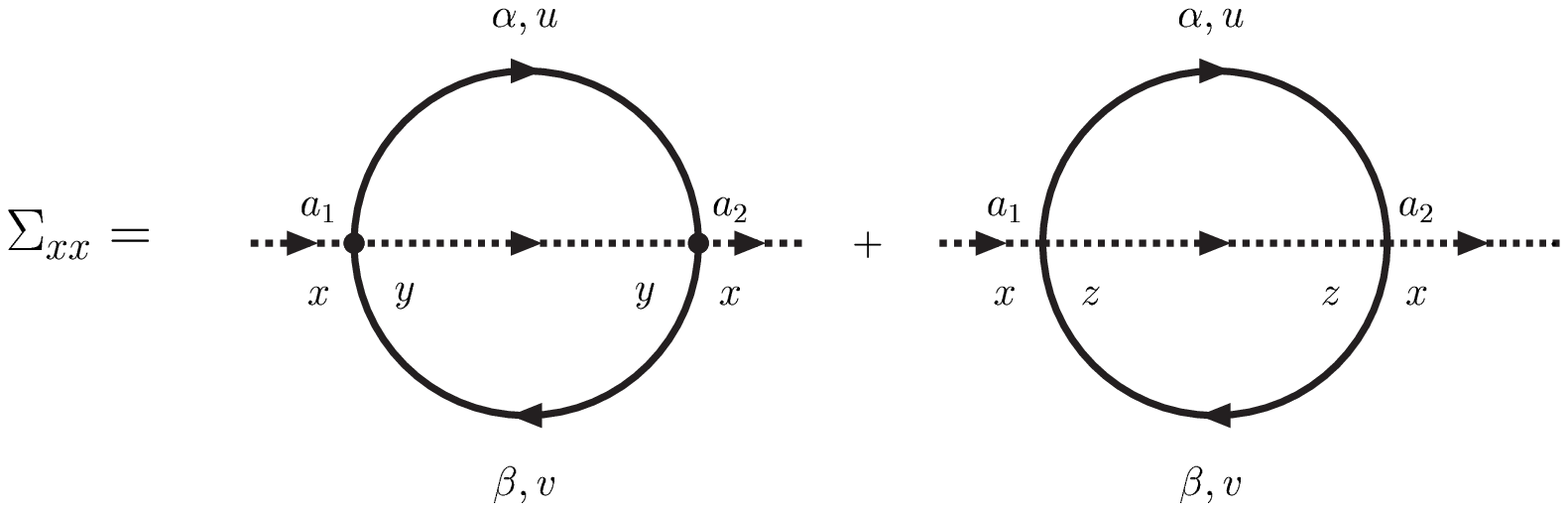}
\]
\caption{\label{Fig.DiagSigmaxxxy}
\sl 
Diagrams of $\Sigma_{xx}$ and $\Sigma_{xy}$.
}
\end{figure}

The first part of the computation reduces the spin and lead indices,
and thus expresses the self energy diagrams as functions of the
``generic'' diagram presented in Fig.~\ref{Fig.DiagSigmaGeneric}, where we allow any potentials $V_{1}$
 and $V_{2}$ for the electrons and any field $B$ for the internal Majorana line.
The main formula is: 
\begin{equation}\label{App.ExprSigmaFromGenericDiag}
\Sigma_{B} (\omega ) = -
\sum_{\empile{\alpha ,\beta \in\{R,L \}}
{\sigma =+,- }}
 \frac{|J_{\alpha \beta }|^{2}}{4}
\biggl (
 f (\omega, B_{s},V_{\alpha },\sigma,V_{\beta },\sigma  ) + 
f (\omega ,0,V_{\alpha },1 ,V_{\beta },-1 )
\biggr).
\end{equation}
In this expression, the Keldysh structure is implicit and by
convention, the Majorana line is an $\eta^{z}$ line when $B_{s}=0$ and an $f$ line otherwise.
To establish (\ref{App.ExprSigmaFromGenericDiag}), we compute the spin
and lead indices of $\Sigma_{xx}$ and $\Sigma_{xy}$ which are given by the diagrams of 
Fig.~\ref{Fig.DiagSigmaxxxy} (with the Feynman rules given in section \ref{KeldyshMethodGal}),
and we use (\ref{DefGB}). 

We now compute the Keldysh structure of the generic diagram $f$.
\begin{figure}[ht]
\[
\figx{6cm}{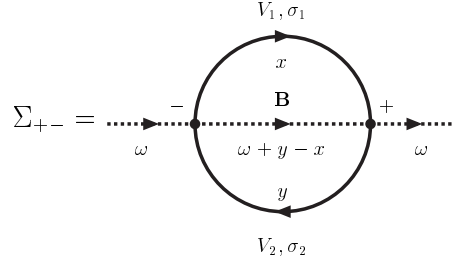}
\]
\caption{\label{Fig.DiagSigmaKeldyshStruct}
\sl 
Computation of the Keldysh structure of the generic diagram.
}
\end{figure}
We have:
\begin{align}\label{App.SigmaPM}
f_{+-} (\omega ) &=  i\pi \int dx dy \,
\rho (x+ \sigma_{1}B_{c} - V_{1}) \rho (y+ \sigma_{2}B_{c} - V_{2})
n_{F} (x-V_{1}) \check n_{F} (y-V_{2}) (1+ {\tilde h} (B)) \delta (\omega
+y -x -B), \nonumber
 \\
f_{-+} (\omega ) &= -  i\pi \int dx dy \,
\rho (x+ \sigma_{1}B_{c} - V_{1}) \rho (y+ \sigma_{2}B_{c} - V_{2})
\check n_{F} (x-V_{1})  n_{F} (y-V_{2}) (1- {\tilde h} (B)) \delta (\omega
+y -x -B),
\end{align}
where ${\check n}_F (x) \equiv n_F (-x)$, and the function ${\tilde h}(\omega)$ is
defined by
\begin{equation}
{\tilde h}(\omega) = \left\{ \begin{array}{l l}
h_B(\omega) & \quad \omega \ne 0, \\
h_z(0) & \quad \omega = 0.
\end{array} \right.
\end{equation}
We then use the relations \cite{RammerSmith}
\begin{align}\label{App.self.RelsigSigpm}
f^{K} &= - \bigl ( f_{+-} + f_{-+}\bigr ),\\
f^A - f^R &= - \bigl ( f_{+-} - f_{-+}\bigr ).
\end{align}
We can perform the integrals in the large bandwidth limit $D\rightarrow \infty$
using 
\[
\int dx \, n_{F} (x+A)\check n_{F} (x) = \frac{A}{e^{A}-1},
\]
and replacing the densities $\rho $ by their finite value $\rho_{0}$
(we can take the limit $D\rightarrow \infty $ under the integral).
Using the definition $V=V_{2}-V_{1}$, we find: 
\begin{align}\label{App.self.sigGen}
\frac{f^K}{2i\pi} &=  \frac{\rho_0^2}{2} (\omega -B +V )
\left(1 - \frac{\widetilde{h} (B)}{\tanh \frac{\omega -B+V}{2}} \right),
\\
\frac{\left( f^A - f^R \right)}{2i\pi} &=  - \frac{\rho_0^2}{2} (\omega -B +V )
\left(\frac{1}{\tanh \frac{\omega -B+V}{2}} - \widetilde{h} (B) \right).
\end{align}
Using (\ref{App.ExprSigmaFromGenericDiag}), and introducing 
\begin{equation}\label{App.Defphi}
\varphi (x)\equiv \frac{x}{\tanh \frac{x}{2}},
\end{equation} 
we now have: 
\begin{align}\label{App.self.sigOrder2}
\nonumber
\sigma_{B} (\omega ) =& \frac{|\rho_0 J_{RL}|^{2}}{4} 
\Bigl [
-h_{B} (B) (\omega -B +V) - h_{z} (0) (\omega +V)
+
\varphi (\omega -B+V) + \varphi (\omega +V)
 + (V\rightarrow -V)
\Bigr ]
\\
& + \sum_{\alpha =R,L} 
 \frac{|\rho_0 J_{\alpha }|^{2}}{4}
\Bigl [
-h_{B} (B) (\omega -B ) - h_{z} (0) (\omega )
+
\varphi (\omega -B) + \varphi (\omega )
\Bigr ],
\\
\nonumber
\sigma_{B}^{K} (\omega ) =& 
- \frac{|\rho_0 J_{RL}|^{2}}{4} 
\left[
 (\omega -B +V) 
\left(
1- \frac{h_{B} (B)}{\tanh \frac{\omega -B +V}{2}}
 \right)
+ 
 (\omega +V) 
\left(
1- \frac{h_{z} (0)}{\tanh \frac{\omega  +V}{2}}
 \right)
 + (V\rightarrow -V)
 \right]
\\
&-  \sum_{\alpha =R,L}
\frac{|\rho_0 J_{\alpha }|^{2}}{4} 
\left[
 (\omega -B ) 
\left(
1- \frac{h_{B} (B)}{\tanh \frac{\omega -B }{2}}
 \right)
+ 
\omega 
\left(
1- \frac{h_{z} (0)}{\tanh \frac{\omega  }{2}}
 \right)
 \right].
\end{align}

Using  the definition of the Majorana Green's function $G^K_{ab}(t)$,
\begin{equation}
G^K_{ab}(t) = -i \left\langle [ \eta_a(t),\eta_b(0) ] \right\rangle,
\end{equation}
we have  $G^K_{ab}(-t) = -G^K_{ba}(t)$.
In the frequency representation, this reads $G^K_{ab}(\omega) = -G^K_{ba}(-\omega)$
and hence we infer that $G^K_{zz}(\omega)$ is odd in frequency. Consequently,
given the form $G^K_{zz} = h_z(\omega) \delta(\omega)$
it is clear that 
\[
h_z(0)=0.
\]
Using this result together with  (\ref{DefhB}) and denoting $x=-h_{B} (B)$, we have:
\[
x= \dfrac{|J_{RL}|^{2} 
\left(
2 x \varphi (V) + 2B \right) + (J_{R}^{2 } + J_{L}^{2})
\left(2x + B  \right)
}
{
|J_{RL}|^{2}\left( 2 \varphi (V) + \varphi (B+V) + \varphi (B-V) \right)
+(J_{R}^{2 } + J_{L}^{2}) \left(2+\varphi (B) \right)
}.
\]
Solving for $x$ and substituting into (\ref{MdotAsAFunctionofHb})
gives Eq. (\ref{MAINRESULT}) of the text.

\bibliography{dot} 

\end{document}